\documentclass[12pt,preprint]{aastex}
\usepackage{subfig}

\usepackage{float}

\usepackage{natbib}
\usepackage{epstopdf}

\usepackage{lineno}

\title{VERITAS Observations of the Nova in V407 Cygni}

\author{
E.~Aliu\altaffilmark{1},
S.~Archambault\altaffilmark{2},
T.~Arlen\altaffilmark{3},
T.~Aune\altaffilmark{4},
M.~Beilicke\altaffilmark{5},
W.~Benbow\altaffilmark{6},
A.~Bouvier\altaffilmark{4},
S.~M.~Bradbury\altaffilmark{7},
J.~H.~Buckley\altaffilmark{5},
V.~Bugaev\altaffilmark{5},
K.~Byrum\altaffilmark{8},
A.~Cannon\altaffilmark{9},
A.~Cesarini\altaffilmark{10},
L.~Ciupik\altaffilmark{11},
E.~Collins-Hughes\altaffilmark{9},
M.~P.~Connolly\altaffilmark{10},
W.~Cui\altaffilmark{12},
G.~Decerprit\altaffilmark{13},
R.~Dickherber\altaffilmark{5},
C.~Duke\altaffilmark{14},
J.~Dumm\altaffilmark{15},
V.~V.~Dwarkadas\altaffilmark{16},
M.~Errando\altaffilmark{1},
A.~Falcone\altaffilmark{17},
Q.~Feng\altaffilmark{12},
J.~P.~Finley\altaffilmark{12},
G.~Finnegan\altaffilmark{18},
L.~Fortson\altaffilmark{15},
A.~Furniss\altaffilmark{4},
N.~Galante\altaffilmark{6},
D.~Gall\altaffilmark{19},
S.~Godambe\altaffilmark{18},
S.~Griffin\altaffilmark{2},
J.~Grube\altaffilmark{11},
G.~Gyuk\altaffilmark{11},
D.~Hanna\altaffilmark{2},
J.~Holder\altaffilmark{20},
H.~Huan\altaffilmark{21},
G.~Hughes\altaffilmark{13},
T.~B.~Humensky\altaffilmark{22},
P.~Kaaret\altaffilmark{19},
N.~Karlsson\altaffilmark{15},
M.~Kertzman\altaffilmark{23},
Y.~Khassen\altaffilmark{9},
D.~Kieda\altaffilmark{18},
H.~Krawczynski\altaffilmark{5},
F.~Krennrich\altaffilmark{24},
M.~J.~Lang\altaffilmark{10},
K.~Lee\altaffilmark{5},
G.~Maier\altaffilmark{13},
P.~Majumdar\altaffilmark{3},
S.~McArthur\altaffilmark{5},
A.~McCann\altaffilmark{2},
J.~Millis\altaffilmark{25,25},
P.~Moriarty\altaffilmark{26},
R.~Mukherjee\altaffilmark{1},
P.~D~Nu\~{n}ez\altaffilmark{18},
R.~A.~Ong\altaffilmark{3},
M.~Orr\altaffilmark{24},
A.~N.~Otte\altaffilmark{27},
D.~Pandel\altaffilmark{28},
N.~Park\altaffilmark{21},
J.~S.~Perkins\altaffilmark{29,30},
M.~Pohl\altaffilmark{31,13},
H.~Prokoph\altaffilmark{13},
J.~Quinn\altaffilmark{9},
K.~Ragan\altaffilmark{2},
L.~C.~Reyes\altaffilmark{32},
P.~T.~Reynolds\altaffilmark{33},
E.~Roache\altaffilmark{6},
H.~J.~Rose\altaffilmark{7},
J.~Ruppel\altaffilmark{31,13},
D.~B.~Saxon\altaffilmark{20},
M.~Schroedter\altaffilmark{6},
G.~H.~Sembroski\altaffilmark{12},
C.~Skole\altaffilmark{13},
A.~W.~Smith\altaffilmark{18},
D.~Staszak\altaffilmark{2},
I.~Telezhinsky\altaffilmark{31,13},
G.~Te\v{s}i\'{c}\altaffilmark{2},
M.~Theiling\altaffilmark{12},
S.~Thibadeau\altaffilmark{5},
K.~Tsurusaki\altaffilmark{19},
J.~Tyler\altaffilmark{2},
A.~Varlotta\altaffilmark{12},
S.~Vincent\altaffilmark{18},
M.~Vivier\altaffilmark{20},
S.~P.~Wakely\altaffilmark{21},
J.~E.~Ward\altaffilmark{9},
T.~C.~Weekes\altaffilmark{6},
A.~Weinstein\altaffilmark{24},
T.~Weisgarber\altaffilmark{21},
R.~Welsing\altaffilmark{13},
D.~A.~Williams\altaffilmark{4},
B.~Zitzer\altaffilmark{12}
}

\altaffiltext{1}{Department of Physics and Astronomy, Barnard College, Columbia University, NY 10027, USA}
\altaffiltext{2}{Physics Department, McGill University, Montreal, QC H3A 2T8, Canada}
\altaffiltext{3}{Department of Physics and Astronomy, University of California, Los Angeles, CA 90095, USA}
\altaffiltext{4}{Santa Cruz Institute for Particle Physics and Department of Physics, University of California, Santa Cruz, CA 95064, USA}
\altaffiltext{5}{Department of Physics, Washington University, St. Louis, MO 63130, USA}
\altaffiltext{6}{Fred Lawrence Whipple Observatory, Harvard-Smithsonian Center for Astrophysics, Amado, AZ 85645, USA}
\altaffiltext{7}{School of Physics and Astronomy, University of Leeds, Leeds, LS2 9JT, UK}
\altaffiltext{8}{Argonne National Laboratory, 9700 S. Cass Avenue, Argonne, IL 60439, USA}
\altaffiltext{9}{School of Physics, University College Dublin, Belfield, Dublin 4, Ireland}
\altaffiltext{10}{School of Physics, National University of Ireland Galway, University Road, Galway, Ireland}
\altaffiltext{11}{Astronomy Department, Adler Planetarium and Astronomy Museum, Chicago, IL 60605, USA}
\altaffiltext{12}{Department of Physics, Purdue University, West Lafayette, IN 47907, USA }
\altaffiltext{13}{DESY, Platanenallee 6, 15738 Zeuthen, Germany}
\altaffiltext{14}{Department of Physics, Grinnell College, Grinnell, IA 50112-1690, USA}
\altaffiltext{15}{School of Physics and Astronomy, University of Minnesota, Minneapolis, MN 55455, USA}
\altaffiltext{16}{Department of Astronomy and Astrophysics, University of Chicago, Chicago, IL, 60637}
\altaffiltext{17}{Department of Astronomy and Astrophysics, 525 Davey Lab, Pennsylvania State University, University Park, PA 16802, USA}
\altaffiltext{18}{Department of Physics and Astronomy, University of Utah, Salt Lake City, UT 84112, USA}
\altaffiltext{19}{Department of Physics and Astronomy, University of Iowa, Van Allen Hall, Iowa City, IA 52242, USA; daniel-d-gall@uiowa.edu, kazuma-tsurusaki@uiowa.edu}
\altaffiltext{20}{Department of Physics and Astronomy and the Bartol Research Institute, University of Delaware, Newark, DE 19716, USA}
\altaffiltext{21}{Enrico Fermi Institute, University of Chicago, Chicago, IL 60637, USA}
\altaffiltext{22}{Physics Department, Columbia University, New York, NY 10027, USA}
\altaffiltext{23}{Department of Physics and Astronomy, DePauw University, Greencastle, IN 46135-0037, USA}
\altaffiltext{24}{Department of Physics and Astronomy, Iowa State University, Ames, IA 50011, USA}
\altaffiltext{25}{Department of Physics, Anderson University, 1100 East 5th Street, Anderson, IN 46012}
\altaffiltext{26}{Department of Life and Physical Sciences, Galway-Mayo Institute of Technology, Dublin Road, Galway, Ireland}
\altaffiltext{27}{School of Physics \& Center for Relativistic Astrophysics, Georgia Institute of Technology, 837 State Street NW, Atlanta, GA 30332-0430}
\altaffiltext{28}{Department of Physics, Grand Valley State University, Allendale, MI 49401, USA}
\altaffiltext{29}{CRESST and Astroparticle Physics Laboratory NASA/GSFC, Greenbelt, MD 20771, USA.}
\altaffiltext{30}{University of Maryland, Baltimore County, 1000 Hilltop Circle, Baltimore, MD 21250, USA.}
\altaffiltext{31}{Institut f\"ur Physik und Astronomie, Universit\"at Potsdam, 14476 Potsdam-Golm,Germany}
\altaffiltext{32}{Physics Department, California Polytechnic State University, San Luis Obispo, CA 94307, USA}
\altaffiltext{33}{Department of Applied Physics and Instrumentation, Cork Institute of Technology, Bishopstown, Cork, Ireland}
\shortauthors{Aliu et al.}

\begin{document}

\begin{abstract}

We report on very high energy (E $>$ 100 GeV) gamma-ray observations of V407 Cygni, a symbiotic binary that underwent a nova outburst producing 0.1--10~GeV gamma rays during 2010 March 10--26.  Observations were made with the Very Energetic Radiation Imaging Telescope Array System during 2010 March 19--26 at relatively large zenith angles, due to the position of V407 Cyg.  An improved reconstruction technique for large zenith angle observations is presented and used to analyze the data.  We do not detect V407 Cygni and place a differential upper limit on the flux at 1.6~TeV of $2.3 \times 10^{-12} \rm \, erg \, cm^{-2} \, s^{-1}$ (at the 95\% confidence level).  When considered jointly with data from {\it Fermi}-LAT, this result places limits on the acceleration of very high energy particles in the nova.

\end{abstract}

\keywords{(Stars:) white dwarfs---(Stars:) novae, cataclysmic variables---Gamma rays: general}

\section{Introduction}\label{sec:Introduction}

In March of 2010, the {\it Fermi}-LAT Collaboration announced a new GeV transient in the galactic plane, FGL J2102+4542, that was identified as a nova outburst in the symbiotic binary V407 Cygni (hereafter, V407 Cyg).  At least seven GeV transients located near the Galactic plane have been discovered by  EGRET, {\it Fermi}-LAT and AGILE.  Only two have been identified at other wavelengths:  V407 Cyg, which is the first nova to be detected at GeV energies, and J0109+6134, which was likely a background blazar \citep{Abdo10,Vandenbrouke10}.  The physical nature of the other five sources is unknown \citep{Abdo10, Chaty10, Hays10, Sabatini10, Tavani97, Vandenbrouke10}, and some of these GeV transients may represent a new class of gamma-ray emitting objects.

The {\it Fermi}-LAT Collaboration reported variable gamma-ray emission in the 0.1--10~GeV band from FGL J2102+4542 during 2010 March 10--26 (MJD 55265--55281) \citep{ATel2487}.  Its flux in gamma rays, binned on a day-to-day basis, peaked 2010 March 13--14 with a flux of $9 \times 10^{-7} \rm \, photons \, cm^{-2} \, s^{-1}$ above 100~MeV \citep{Abdo10}.  The GeV gamma-ray activity lasted approximately two weeks.  The initial report of GeV emission triggered Very Energetic Radiation Imaging Telescope Array System (VERITAS) observations of the object at very high energy (VHE; E $>$ 100GeV) as part of an ongoing campaign to observe transients detected by {\it Fermi}-LAT.

Using multi-wavelength data, it was determined that the new transient was most likely associated with V407 Cyg, a binary system consisting of a Mira-type pulsating red giant and a white dwarf companion \citep{Abdo10}.  A nova outburst from V407 Cyg was detected in the optical waveband on 2010 March 10 \citep{optical} with a  magnitude of $\sim6.9$, while pre-outburst magnitudes from the previous two years of monitoring ranged between magnitude 9 and 12 \citep{Abdo10}. V407 Cyg has been optically monitored for decades and has experienced previous outbursts, but the system had never been observed to be as bright as during the nova \citep[e.g.,][]{orbit, monitoring}.  The onset of the optical outburst corresponds to the first significant detection of the source by the {\it Fermi}-LAT on 2010 March 10.

Novae in red giant/white dwarf systems have been known to produce expanding shocks that can result in X-ray emission \citep[e.g., the recurrent nova RS Oph,][]{sokoloski, bode}, and indeed, X-ray emission from V407 Cyg was detected after the nova \citep{Abdo10, tommy}.  Based on the observed X-ray emission from the 2006 nova outburst of RS Oph, before the launch of {\it Fermi}-LAT, \citet{rsoph} suggested that particles could be accelerated in novae up to TeV energies, but gamma-ray emission from a nova had never previously been detected.  Here, we discuss the VERITAS observations of V407 Cyg and their implications for gamma-ray emission from the nova.  We also describe an improved event reconstruction technique for stereo observations by imaging atmospheric Cherenkov telescopes (IACTs) made at large zenith angles (LZA).

\section{Observation and Analysis}
\subsection{VERITAS Observations}

VERITAS is a ground-based VHE gamma-ray observatory located at the Fred Lawrence Whipple Observatory in southern Arizona. It consists of four IACTs sensitive from approximately 100 GeV to above 30 TeV. Each VERITAS telescope has a 12~m tessellated reflector with a total area of 110\,m$^{2}$. Each camera's focal plane contains 499 closed-packed circular photomultiplier tubes, giving a total field-of-view of $3.5^\circ$.  Gamma-rays incident onto the upper atmosphere induce a particle cascade, called an air shower, in which some charged particles have sufficient speed to emit Cherenkov light.  The direction and energy of the original gamma ray can be reconstructed from images of the Cherenkov light recorded by the telescopes. When observing at small zenith angles ($< 40^{\circ}$), the array has an energy resolution of $15\%$ at 1 TeV and an angular resolution of better than $0.1^\circ$ at 1 TeV \citep{VERITAS}. For observations at LZA, the energy and angular resolution are degraded and the energy threshold is increased.

VERITAS observed V407 Cyg for several nights after the announcement of the {\it Fermi}-LAT detection, during days 9--16 of the outburst (2010 March 19--26).  The zenith angle of these observations ranged between $50^\circ$ and $66^\circ$.  The VERITAS telescopes are regularly operated in a mode called wobble mode, during which the location of the object to be observed is offset from the center of the field of view (FoV) by 0.5$^{\circ}$, allowing for simultaneous source and background measurements \citep{Fomin}.  The offset direction cycles between north, south, east and west for sequential observing segments to reduce systematic effects.  After filtering the data for contamination due to poor weather or instrumental problems, 304 minutes of live time remained from the original 335 minutes of observations, see Table~1.

To test the improved reconstruction technique discussed in Section \ref{sec:Two Reconstruction Methods}, VERITAS observations of the Crab Nebula were also analyzed. We selected 203 minutes of good time intervals from 17 data segments taken on the Crab Nebula during 2010 March 12--16 (MJD 55267--55271) with similar zenith angles ranging from $55^\circ$  to $65^\circ$.  All data were analyzed using the standard analysis package for VERITAS data \citep{vegas}.

\subsection{Event Reconstruction}  \label{sec:Two Reconstruction Methods}

The raw data were calibrated and cleaned, and quality selection criteria based on the number of photomultiplier tubes contained in the images and the position of the image in the camera were applied.  The shape and orientation of the gamma-ray images were parametrized by their principal moments \citep{Hillas85}.  In order to produce gamma-ray images of the sky, it is necessary to reconstruct the putative source location for each shower in the camera plane (hereafter ``arrival direction").  When imaging showers with multiple IACTs, the arrival direction of a shower is usually found using simple geometric arguments.  The major axes of the images produced by a shower in each IACT camera intersect near the location of the arrival direction.  The shower arrival direction is calculated by minimizing the perpendicular distance to each image's semi-major axis, weighted by the size of each image.  This method, here called the standard method, is effective at small zenith angles.  However, at LZA, the major axes of the air shower images from an individual gamma-ray event are generally close to parallel.  Thus, the uncertainty of the intersecting point increases, resulting in a loss of angular resolution.  Due to this effect, a reconstruction technique that does not depend on the intersection of the axes is desirable for LZA observations. 

The displacement method is a direction reconstruction algorithm that is useful for LZA observations \citep{dispicrc}.  In very general terms, it consists of calculating the arrival direction using the shape and brightness of a given air shower image. More specifically, it relies on the determination of the \emph{disp} parameter, defined as the angular distance from the image centroid to the arrival direction. This method was used by several experiments in the past~\citep{lessard01, kranich03, domingo05}, with varying ways of calculating \emph{disp}.

The basis of the displacement method is the relationship of the \emph{disp} parameter to other image parameters~\citep{hofmann99}.  The implementation of the algorithm in VERITAS is as follows: we estimate \emph{disp} as a function of three other image parameters, \emph{size}, \emph{length} and \emph{width}~\citep{Hillas85}, using Monte Carlo simulated gamma-ray showers. The method results in two different arrival directions, one on each side of a telescope image along the semi-major axis, also known as head-tail ambiguity~\citep{hofmann99}. This ambiguity is eliminated by choosing the cluster of arrival directions closest to one another, one coming from each image. Finally, the arrival direction is estimated independently for each telescope image and an average weighted by {\it size} is taken. This method proves to be more powerful than the standard method~\citep{vegas} when reconstructing events with zenith angle larger than $50^\circ$. Quantitatively, an improvement of $\sim$30\% in detection significance for a source having 1\% of the strength of the Crab Nebula has been observed.

\subsection{Event Selection} \label{sec:Event Selection}

The cosmic ray background rate for IACTs is typically more than $10^{4}$ times the gamma ray rate, so it is important to reduce this background while retaining as many gamma-ray events as possible. By exploiting the differences in the development of gamma ray and cosmic ray induced showers, the background due to cosmic rays can be reduced significantly, while still retaining a high fraction of gamma-ray like events.  The background reduction is performed by placing standard selection criteria, optimized using Monte Carlo simulations and real data from the Crab Nebula on the shower image parameters. The selection criteria for the size of the telescope images, the mean scaled width and mean scaled length parameters \citep{Daum97,msw}, the height of maximum Cherenkov emission and the angular distance from the anticipated source location to the reconstructed arrival direction of each shower ($\theta$) are given in Table \ref{cuts}.

To perform a background subtraction of the surviving cosmic ray events, an estimation of these background counts is made using the reflected-region background model \citep{refreg}.  Events within an angular distance $\theta$ of the anticipated source location are considered ON events.  Background measurements (OFF events) are taken from regions of the same size and at the same angular distance from the center of the FoV.  For this analysis, a minimum of eight background regions was used.  The excess number of events from the anticipated source location is found by subtracting the number of OFF events (scaled by the relative exposure, $\alpha$) from the ON events.  Statistical significances are calculated using a modified version of Eqn. 17 of \citet{LiMa} to
allow for varying number of off-source regions due to nearby star 60 Cygni \citep{modlima}.  More details about VERITAS, the calibration procedure and the analysis techniques can be found in \citet{VERITAS}.

\section{Results} \label{sec:Results}

Analysis of the VERITAS data did not show a significant detection at the location of V407 Cyg.  The results from both event reconstruction methods were used to calculate upper limits on the flux from V407 Cyg with the method described by \citet{Rolke} and the assumption of a Gaussian-distributed background.  

The upper limits for V407 Cyg are calculated at the decorrelation energies of 1.8 TeV for the standard method and 1.6 TeV for the displacement method and assume that any emission takes the form of a power law with a photon index of -2.5.  The decorrelation energy is the energy at which the dependence of the upper limit calculation on the assumed photon index is minimized.  This energy is found by performing multiple upper limit calculations, with different spectral indices, and determining the region where the resulting upper limit functions intersect.  The energy threshold for the observations of V407 Cygni with VERITAS, defined as the maximum of the product of the assumed spectral shape and the effective area, is 1.2 TeV for both methods.

\begin{figure}[ht]
\centering
    \subfloat[standard method]{\label{fig:Fermistandard}\includegraphics[scale=0.4]{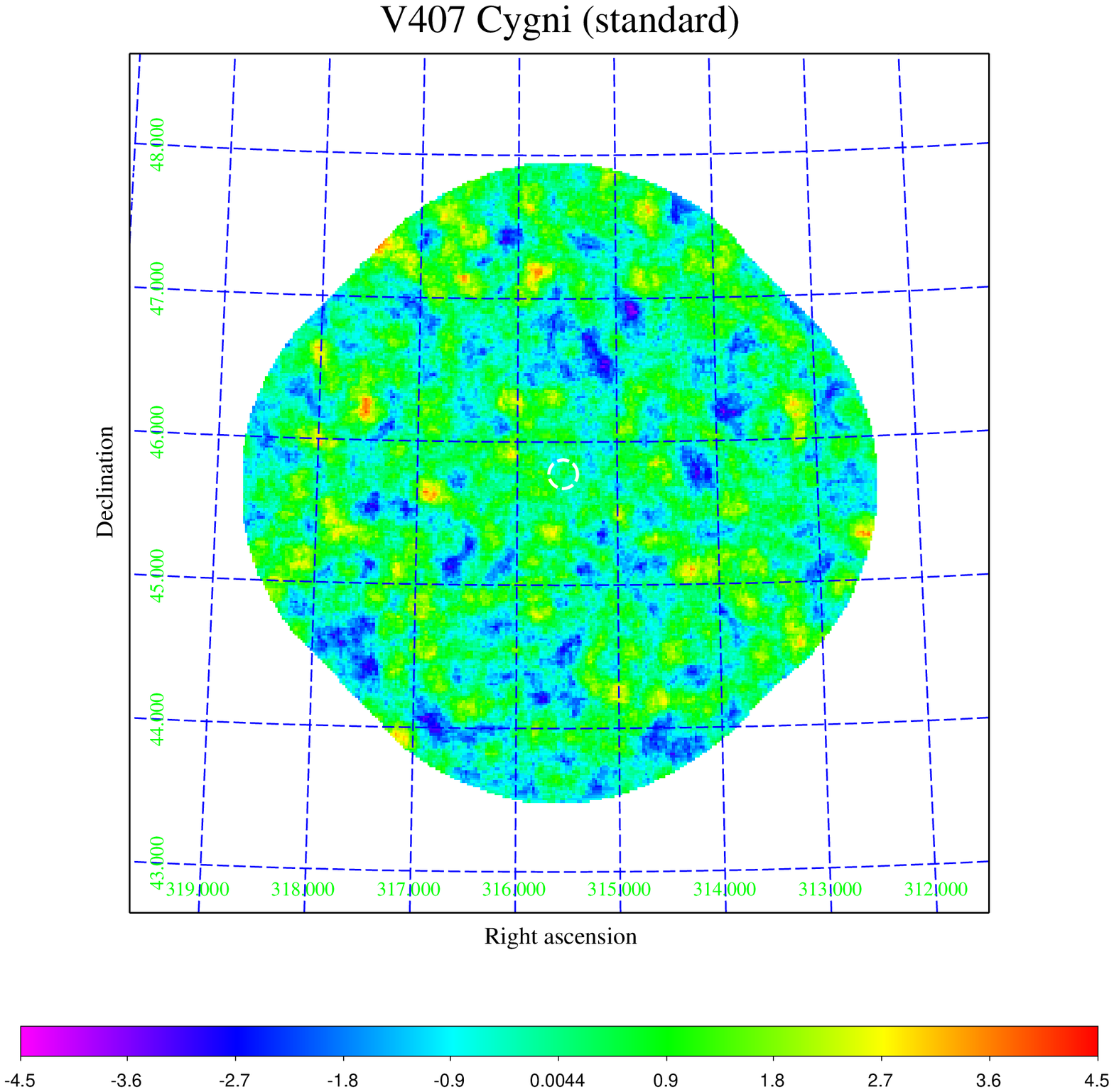}}
\hspace{12px}
    \subfloat[displacement method]{\label{fig:Fermilza}\includegraphics[scale=0.4]{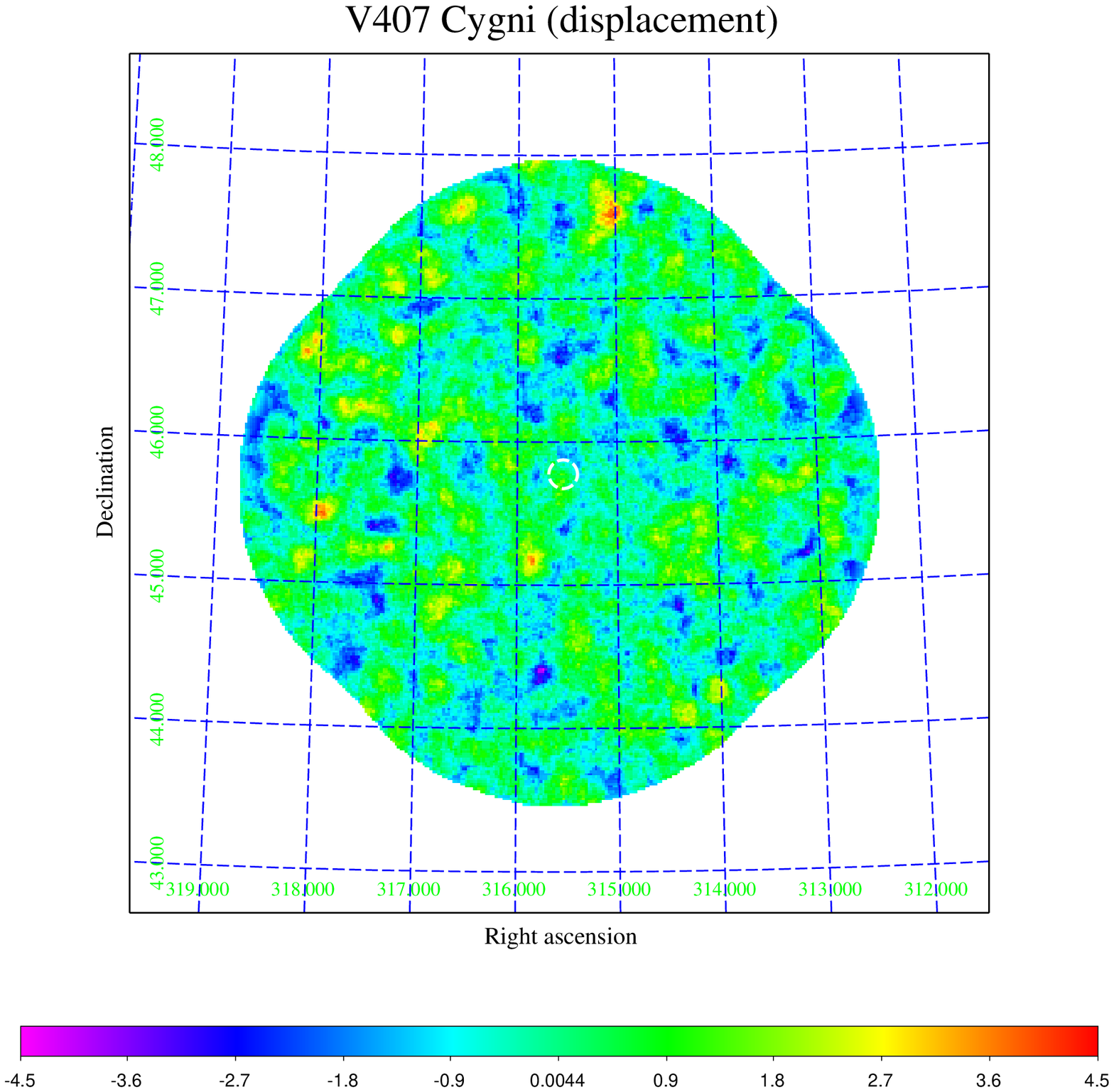}}
    \caption{VERITAS significance maps of V407 Cyg reconstructed by the two methods. The white circle at the center indicates the location of the {\it Fermi}-LAT detection and the size of the source region used for the VERITAS analysis.}
    \label{fig:mapfermi}

\end{figure}

\begin{figure}[ht]
  \centering
 \subfloat[standard method]{\includegraphics[scale=0.4]{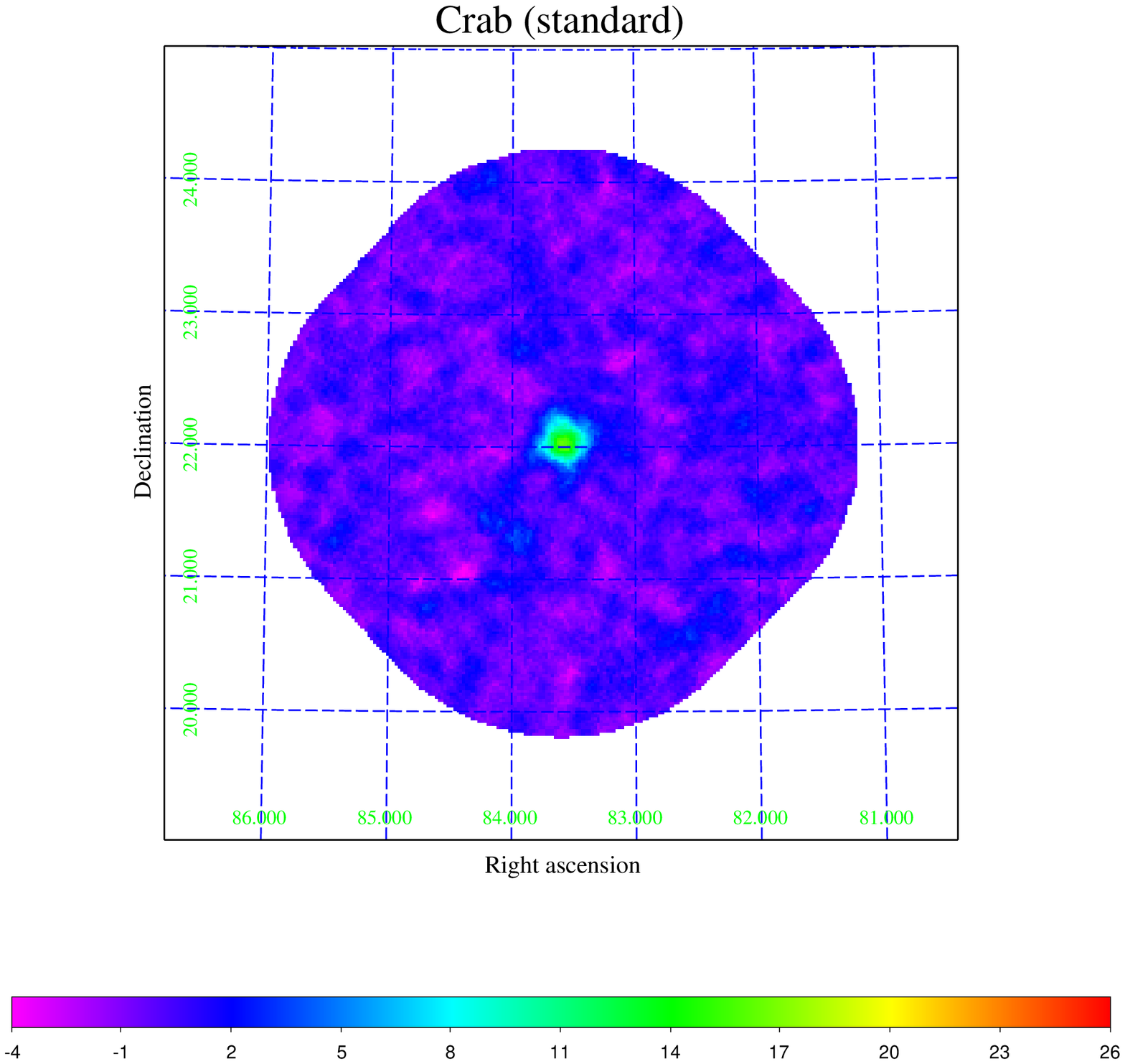}} 
       \hspace{12pt}              
 \subfloat[displacement method]{\includegraphics[scale=0.4]{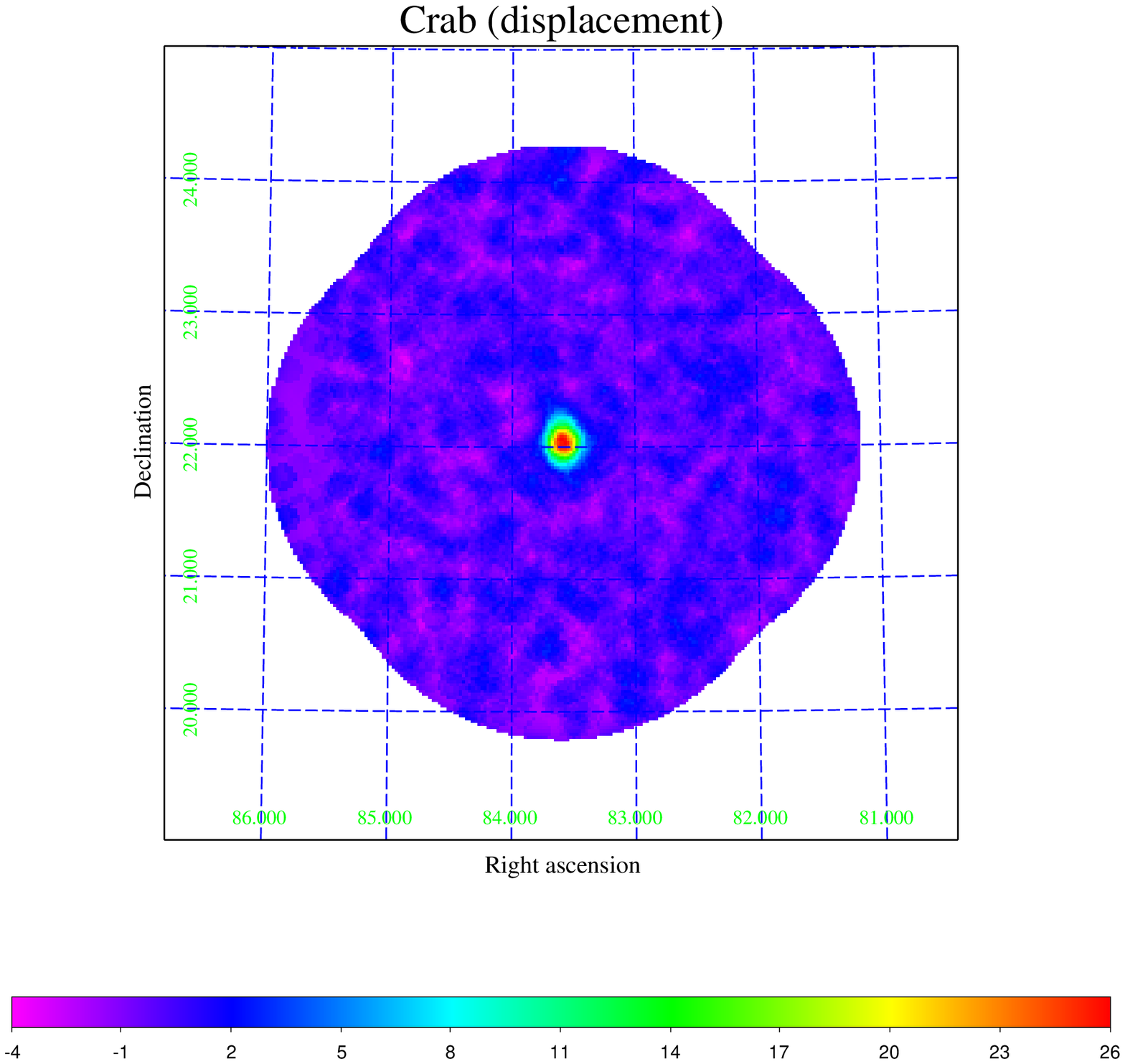}}
  \caption{VERITAS significance maps of the Crab Nebula reconstructed by the two methods.}
  \label{fig:mapcrab}
\end{figure}

The analysis results for V407 Cyg are presented in Table \ref{results} and Figure \ref{fig:mapfermi}.  In addition, results from observations of the Crab Nebula at similar zenith angles are presented in Table \ref{results} and Figure \ref{fig:mapcrab}.  The efficiency of the displacement method for event reconstruction can be observed in the increase in both gamma-ray rate and significance for the Crab measurements.  The increased sensitivity also results in a reduction of the upper limit for V407 Cyg compared to the standard method.

\section{Discussion} \label{sec:Discussion}

The GeV detection of V407 Cyg provides evidence for previously unobserved gamma-ray emission from novae in white dwarf/red giant systems.  Expanding shock waves have been known to accelerate particles to high energies, and gamma rays are observed from supernova remnants. The discovery by the {\it Fermi}-LAT team, however, suggests that the same phenomenon occurs in some novae, adding a new class of gamma-ray emitting objects.  The lack of a significant detection in the VHE band suggests that either particles were not accelerated to sufficient energies to produce VHE photons during the V407 Cyg outburst or that VHE photons were produced, but then absorbed.

The key to creating gamma rays is the acceleration of sufficiently energetic charged particles. In the case of V407 Cyg, the expanding matter from the nova collides with the stellar wind from the red giant and causes a shock, which accelerates the particles near the shock to relativistic energies. A rough estimation for the maximum energy attainable by first order Fermi acceleration of a particle at a shock can be found, following the discussion of \citet{Longair}. If $B$ is the magnetic flux density where a shock proceeds and the shock travels with velocity $U$, the maximum energy of a particle with charge $Ze$ is $E_{max} = \int ZeBU^2 dt$, where $t$ is the time allowed for particle acceleration. This means that the highest attainable energy is proportional to the magnetic field in which the nova travels, the square of shock speed and the time for acceleration. The mean magnetic field in the shock can be estimated as $
B=[32\pi \rho(R) kT]^\frac{1}{2}$, where $\rho(R)$ is the density of gas molecules with respect to the distance from the center of the red giant and T is the wind temperature \citep{Abdo10}.  We assume a wind temperature of T=700K, corresponding to the temperature of the dust envelope measured by \citet{orbit}.  \citet{windtemp} were able to directly measure the temperature of the red giant wind in a similar symbiotic system, EG And, and found that it can reach $\sim$8000 K near the red giant.  Using this wind temperature would increase the estimate for maximum particle energy by a factor of three.

\citet{Orlando11} carried out detailed hydrodynamic simulations of the V407 Cyg nova with various gas distribution models and could accurately reproduce the X-ray light curve of V407 Cygni. Their model for the distribution of gas that best reproduced the light curve included what they call a ``circumbinary density enhancement," a region of density exceeding the typical $R^{-2}$ profile of the stellar wind in the binary system, and had a binary separation of 15.5 AU \citep{Orlando11}.  For the temporal profile of the nova shock velocity, we used the equation that \citet{halpha} found from fits to the broad components of the H$\alpha$ spectra they measured beginning at day 2.3 after the outburst (13 March 2010) and thereafter: $U = 4320 - 5440 \log t + 2635(\log t)^2 - 460 (\log t)^3$.  For the velocity between day 0 and day 2 of the outburst, we can assume two cases that bound the possible velocity profiles: i) the nova shell experienced free expansion at a constant velocity before day 2, with the assumption that the mass collected by the nova shell during this period was small (free expansion model) or ii) extrapolate the above equation for the velocity to times before day 2 (extrapolation model).  We then find that at the start of VERITAS observations of V407 Cyg (day 9 of the outburst), $E_{max} \sim 1.4$ TeV for the free expansion model and $\sim 3.0$ TeV for the extrapolation model.  This suggests that particles could have been accelerated to TeV energies by the time of the VERITAS observations.

To check the importance of absorption, we calculated the opacity along the photon path for gamma rays generated at the shock front. Electron-positron pair production via photon-photon collision is the dominant interaction \citep{Gould}.  We modeled the red giant spectrum as a blackbody with a temperature of 2500~K and a radius of 500~$R_{\odot}$, found the photon density as a function of position and energy following \citet{Nunez}, and used the cross section for the photon-photon collision from \citet{Gould}.  We found that the opacity for TeV photons only becomes significant when the TeV emission region is located directly behind the red giant with the system viewed edge on.  Though this case cannot be ruled out, it is statistically unlikely.  In addition, if the suggested orbital parameters of \citet{orbit} are accurate, the system is unlikely to have been in such an orientation at the time of the nova.

The upper limits placed by VERITAS can put some restrictions on the gamma-ray emission mechanism in V407 Cyg.  Two physical models of gamma-ray production at the shock-front have been suggested \citep{Abdo10}. In the hadronic model, gamma rays are produced in the decay of $\pi^{0}$ particles generated by collisions of high energy protons accelerated in the shock.  In the leptonic model, gamma rays are produced via inverse-Compton scattering of infrared photons emitted from the red giant on high energy electrons accelerated in the shock.  

The electron threshold energy for production of a gamma-ray photon via inverse-Compton scattering off the red giant photons can be estimated as: $E> (E_\gamma/2) \,\left(1+\sqrt{1+{{m_e^2\,c^4}\over {E_\gamma \, \epsilon}}}\right)$, where $E$ is the electron threshold energy, $E_{\gamma}$ is the gamma-ray energy and $\epsilon$ is the energy of the red giant photons.  The electron threshold energy for a 1~TeV gamma ray scattering off 0.6 eV photons at the peak of the red giant spectrum is 1.1~TeV.  Though the above calculation indicates that particles could reach TeV energies if continuously accelerated for the full nine days from the initial outburst to the start of the VERITAS observations, the inverse-Compton cooling time would be significantly less than a day (the time estimated by \citet{Abdo10} for 5 GeV electrons), meaning electrons that are accelerated in the first few days of the outburst would not likely retain sufficient energy to produce VHE photons by the time of the VERITAS observations.  VHE emission near the time of the VERITAS observations would therefore require freshly accelerated particles, however, recently accelerated particles would likely not have enough time to reach TeV energies.  If electrons did reach TeV energies, they would be approaching the Klein-Nishina regime, where the cross section for inverse-Compton scattering would be reduced, resulting in a longer cooling time.  However, electrons with TeV energies would be well above the exponential cutoff, $3.2^{+2.6}_{-0.1}$~GeV, of the electron spectrum in the best-fit leptonic model of \citet{Abdo10}.  These two factors imply that the VERITAS upper limits place no new constraints on leptonic models.  

\begin{figure}[ht!]
  \includegraphics {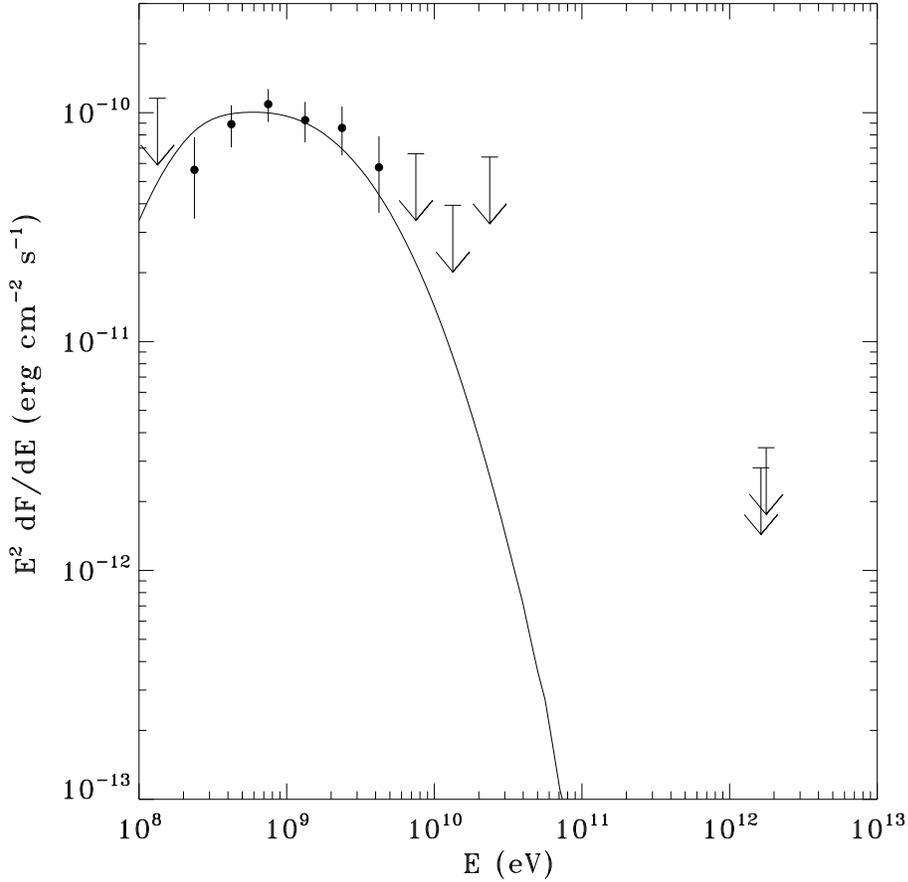}
  \caption{Spectrum of V407 Cyg (FGL J2102+4542) measured by the {\it Fermi}-LAT \citep{Abdo10} and VERITAS upper limits. Vertical bars indicate 1$\sigma$ statistical error, and arrows indicate 2$\sigma$ upper limit. The rightmost arrows show the 99\% confidence level (3$\sigma$) VERITAS upper limit calculated using the displacement method (at 1.6 TeV) and the standard method (at 1.8 TeV) for event reconstruction (See section \ref{sec:Results}). The fitting curve was constructed with the method of \citet{Kamae} with the parameters ($s_{p}$, log$(E_{cp})$)=(2.15, 1.5) (See section \ref{sec:Discussion}).}
  \label{fig:spectrum}
\end{figure}

For hadronic models, the {\it Fermi}-LAT data alone provide relatively poor constraints on the extension of the proton spectrum to high energies.  In particular, the cutoff energy $(E_{cp})$ is not well bounded from above if the spectral index is steep.  The VERITAS data can be used to improve the constraints on the hadronic model parameters.  To do so, we re-fit the hadronic model used by \citet{Abdo10} to the {\it Fermi}-LAT points with the addition of the VERITAS upper limit.  Figure~\ref{fig:spectrum} shows the {\it Fermi}-LAT data \citep{Abdo10} and the VERITAS flux upper limit compared to the best fitting hadronic model.  The gamma-ray spectrum is calculated via the method of \citet{Kamae}, assuming a cosmic proton spectrum of the following form: $N_{p}=N_{p,0} (W_{p} + m_{p}c^{2})^{-s_{p}}e^{-W_{p}/E_{cp}}$ (protons~GeV$^{-1}$), where ($W_{p}, E_{cp}, m_{p}$ are kinetic energy, cut-off energy, and mass of the proton and $s_{p}$ is the power law index).

\begin{figure}[ht!]
  \subfloat[{\it Fermi}-LAT]{\includegraphics[width=3in]{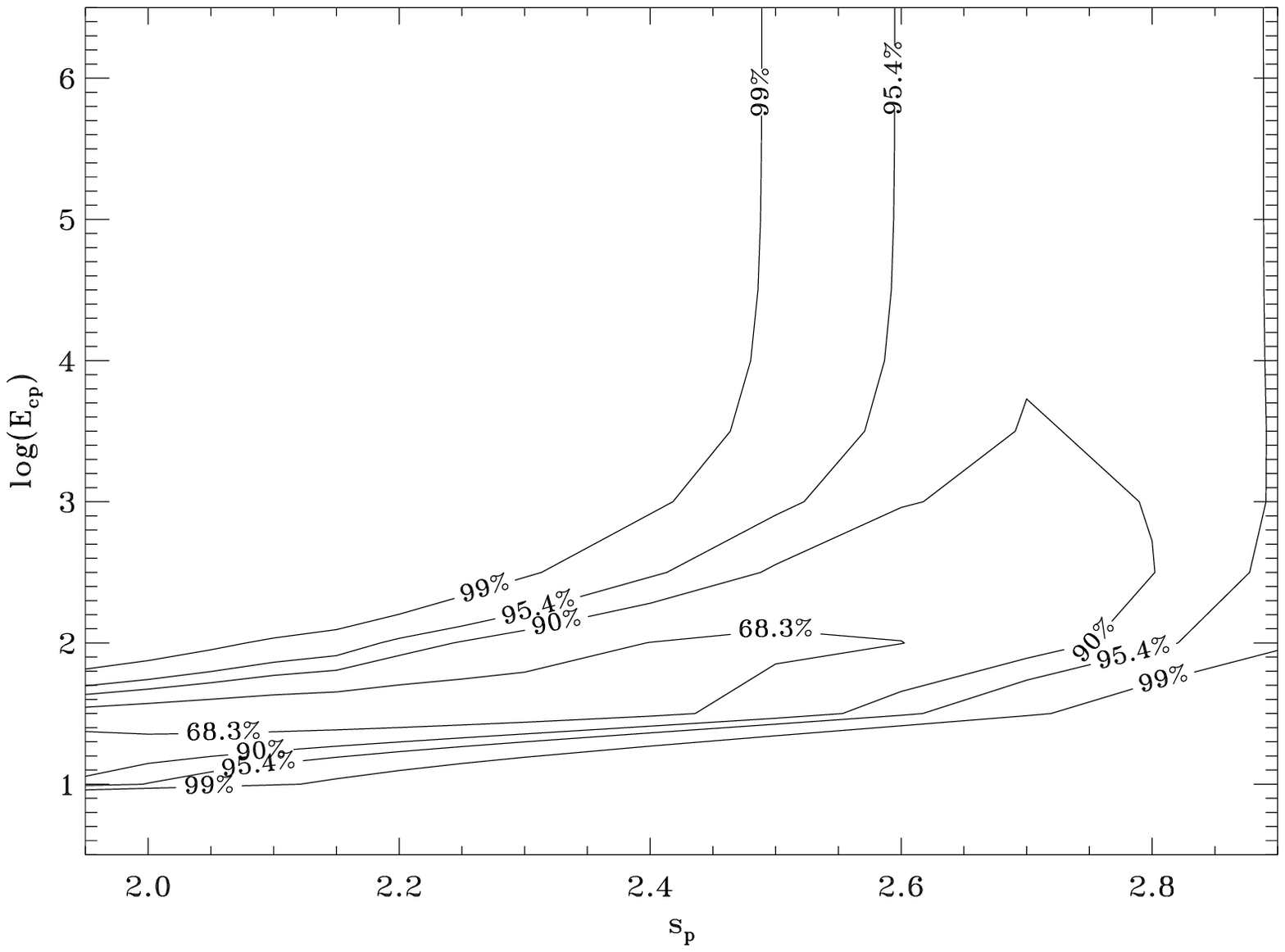}} 
       \hspace{12pt}              
  \subfloat[{\it Fermi}-LAT+VERITAS]{\includegraphics[width=3in]{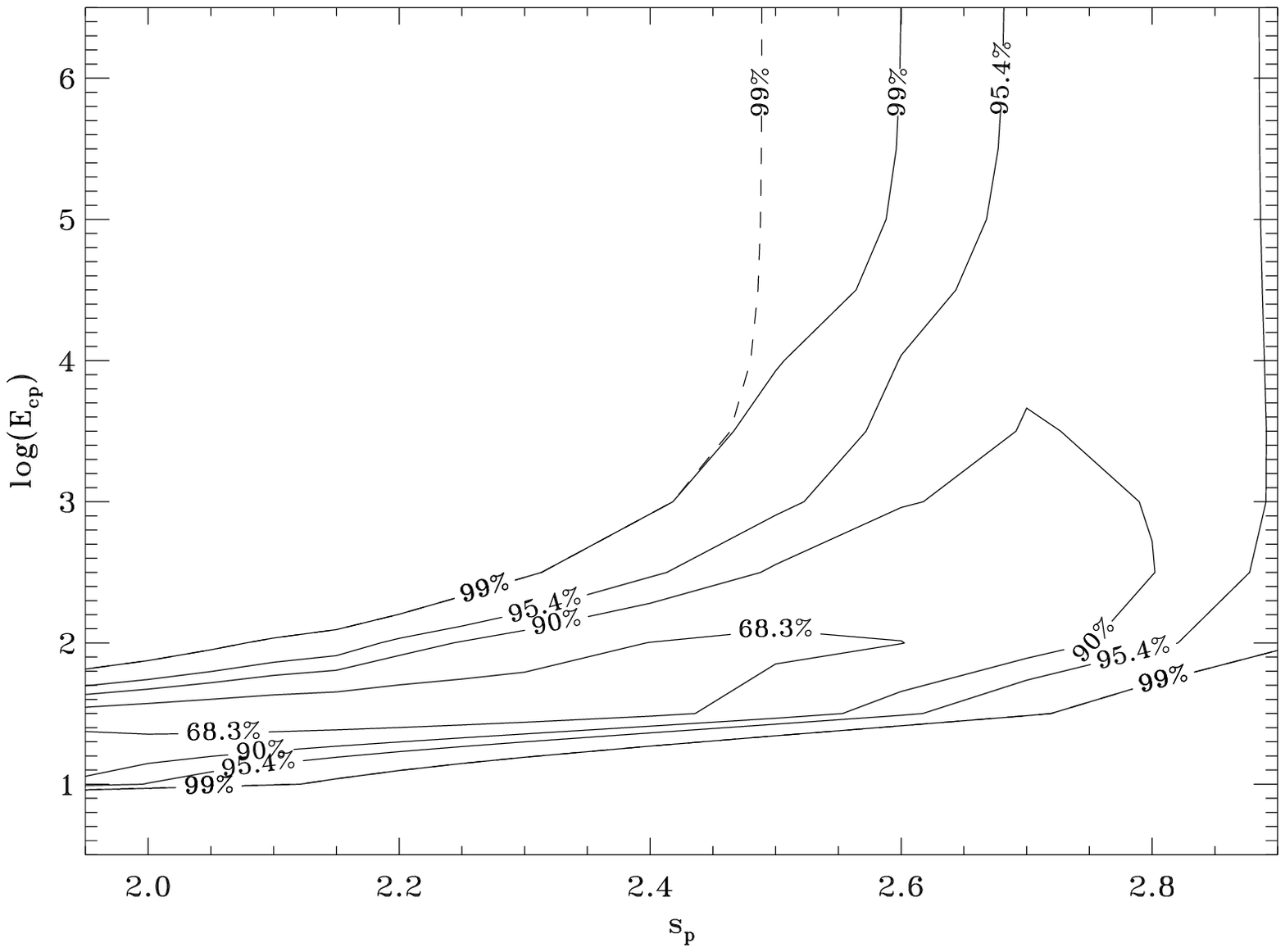}}

  \caption{Confidence region maps for the hadronic model of gamma-ray production in V407 Cyg constrained by {\it Fermi}-LAT data alone (a) and constrained by the {\it Fermi}-LAT data plus the VERITAS upper limit (b).  The x-axis gives the spectral slope ($s_{p}$) and the y-axis is the logarithm of the cutoff energy ($E_{cp}$ in GeV).  The numerical values correspond to the confidence level of the fit with the given values of the parameters.  The 99\% contour from {\it Fermi}-LAT data alone is shown as a dashed line in (b) for comparison.}
  \label{fig:revisedmap}
\end{figure}

Figure~\ref{fig:revisedmap} shows a confidence region map for the parameters of the hadronic model using both the {\it Fermi}-LAT and VERITAS data.  The gamma-ray spectrum was modeled as described above and fitted to the {\it Fermi}-LAT data by varying $s_{p}$ and $E_{cp}$.  The spectrum was then compared to the VERITAS upper limit, and a $\chi^{2}$ value for the VERITAS data point alone was calculated.  Specifically, we calculated the model flux in the VERITAS energy band and compared this to the flux upper limit determined via the displacement method.  This $\chi^{2}$ value was then added to the $\chi^{2}$ calculated for the {\it Fermi}-LAT data.  The confidence levels were then calculated for the two parameters of interest, $s_{p}$ and $E_{cp}$.  As can be seen from Figure~\ref{fig:revisedmap}, the VERITAS observations place greater restrictions on the model proton spectral index for high cutoff energies.  The 90\% confidence limits are $E_{cp}\lesssim$ 5 TeV (comparable to $E_{max}=3$ TeV, calculated above for the extrapolation model) and $E_{cp}\gtrsim$ 0.01 TeV (much lower than $E_{max}$).  It is possible that the peak energy of the particles produced by the shock could be reduced if the magnetic field is weaker than estimated above.  \citet{tommy} argue for a larger binary separation of 20--25 AU, based on the presence of lithium burning in the Mira, and evidence that the white dwarf in the system is massive.  Using this larger separation distance would lead to weaker magnetic fields.  The limits placed by the VERITAS observations are near the threshold for the observations, so it is also possible that simply not enough particles were accelerated to high enough energies to produce a significant detection by VERITAS.

\begin{acknowledgments}

{\it Acknowledgments.} We would like to thank Pierre Jean of the {\it Fermi}-LAT team for providing the contour data from the {\it Fermi}-LAT results and useful discussion. This research is supported by grants from the U.S. Department of 
Energy Office of Science, the U.S. National Science Foundation and the Smithsonian Institution, by NSERC in 
Canada, by Science Foundation Ireland (SFI 10/RFP/AST2748) and by STFC in the U.K.  We acknowledge the excellent 
work of the technical support staff at the Fred Lawrence Whipple Observatory and at the collaborating 
institutions in the construction and operation of the instrument. \end{acknowledgments}

\begin{table}[ht]
  \centering
  \small
  \begin{center}
        \caption{Summary of the VERITAS data sets presented in this work.}
    \begin{tabular}{|c||c|c|c|c|c|} \hline
	Target & Period & Useful Duration[min] & Zenith Angle[$^\circ$] \\\hline\hline
	V407 Cyg & Mar. 19-26, 2010   & 304 & 50-66 \\\hline
	Crab Nebula & Mar. 12-16, 2010 & 203 & 55-65 \\\hline
    \end{tabular}
  \end{center}
\end{table}

\begin{table}[htdp]
\begin{center}
\caption{Selection criteria used for the VERITAS analysis. For an explanation of these parameters, see Section \ref{sec:Event Selection}.}
\begin{tabular}{|l| c|}
\hline
Parameter & Selection Criteria \\ \hline \hline
Image size & $>$  400 digital counts ($\sim$ 75 photoelectrons)\\ \hline
Mean Scaled Width & 0.05 $<$ MSW $<$ 1.15 \\ \hline
Mean Scaled Length & 0.05 $<$ MSL $<$ 1.3 \\ \hline
Height of Shower Maximum& $>$ 7 km\\ \hline
$\theta$ & $<$ 0.1$^{\circ}$ \\ \hline
\end{tabular}
\label{cuts}
\end{center}
\end{table}

\begin{table}[ht]
  \begin{center}
      \caption{A summary of the VERITAS analysis results.  Flux upper limits for V407 Cyg from VERITAS observations are calculated assuming a photon power-law
index of -2.5 and taken at the decorrelation energy, minimizing the effect of the assumed spectral index.  The displacement method produces both a higher significance for the Crab Nebula data and a more sensitive upper limit for V407 Cyg.}
  \small
    \begin{tabular}{|p{60mm}||c|c|c|c|} \hline
    Reconstruction method & \multicolumn{2}{|c|}{Standard} & \multicolumn{2}{|c|}{Displacement} \\\hline
    Source & Crab & V407 Cyg & Crab & V407 Cyg  \\\hline\hline
    Exposure [min] &203&304&203&304\\\hline
    ON (Source) counts  & 255 & 91 & 300 & 76 \\\hline
    OFF (Background) counts & 744 & 841 & 351 & 630 \\\hline
    $\alpha$ (See Sec.\ref{sec:Event Selection}) & 0.111 & 0.125 & 0.111 & 0.125 \\\hline
    Significance & $15.7\sigma$ & $0.5\sigma$ & $25.1\sigma$ & $1.0\sigma$\\\hline
    Rate [photons min$^{-1}$] & $0.97\pm0.09$ & $0.02\pm0.04$ & $1.40\pm0.09$& $0.03\pm0.03$ \\\hline
    Energy threshold [TeV]&1.5 &1.2 & 1.7 &1.2 \\ \hline
     Decorrelation energy [TeV]&- & 1.8 &- & 1.6 \\ \hline
    Flux upper limit at decorrelation energy ($95\%$ c.l.) [E$^{2}$*$d$F/$d$E; erg \,cm$^{-2}$\,s$^{-1}$]   & -  & $2.7\times10^{-12}$  & - & $2.3\times10^{-12}$ \\\hline
    \end{tabular}
     \label{results}
  \end{center}
\end{table}

\end{document}